\documentclass{emulateapj}

\usepackage{natbib}
\usepackage{comment}
\usepackage{url}

\newcommand{\etal}{et al.}

\newcommand{\ppdot}{$P$-$\dot{P}$~}

\slugcomment{accepted on Jun 30, 2009}

\shorttitle{New Limits on Radio Emission from XDINSs}
\shortauthors{Kondratiev \etal}

\begin{document}

\title{New Limits on Radio Emission from X-ray Dim Isolated Neutron Stars}

\author{V.~I.~Kondratiev\altaffilmark{1,2,4},
M.~A.~McLaughlin\altaffilmark{1,2,3},
D.~R.~Lorimer\altaffilmark{1,2},
M.~Burgay\altaffilmark{5},
A.~Possenti\altaffilmark{5},
R.~Turolla\altaffilmark{6,7}, 
S.~B.~Popov\altaffilmark{8}, and S.~Zane\altaffilmark{7}}

\altaffiltext{1}{Department of Physics, West Virginia University, 210 Hodges Hall, 
	 Morgantown, WV 26506; 
	 vlad.kondratiev@mail.wvu.edu, maura.mclaughlin@mail.wvu.edu, duncan.lorimer@mail.wvu.edu}
\altaffiltext{2}{National Radio Astronomy Observatory, P.O. Box 2, Green Bank, WV 24944}
\altaffiltext{3}{Alfred P. Sloan Research Fellow}
\altaffiltext{4}{Astro Space Center of the Lebedev Physical Institute, Profsoyuznaya str. 84/32, Moscow 117997, Russia}
\altaffiltext{5}{INAF - Osservatorio Astronomico di Cagliari, Loc. Poggio dei Pini, Strada 54, 09012, Capoterra (CA), Italy;
	 burgay@ca.astro.it, possenti@ca.astro.it}
\altaffiltext{6}{University of Padua, Department of Physics, via Marzolo 8, 35131 Padova, Italy;
	 roberto.turolla@pd.infn.it}
\altaffiltext{7}{Mullard Space Science Laboratory, University College London, Holmbury St. Mary, Dorking Surrey, RH5 6NT, UK;
	sz@mssl.ucl.ac.uk}
\altaffiltext{8}{Sternberg Astronomical Institute, Universitetski pr. 13, Moscow 119992, Russia;
	sergepolar@gmail.com}

\begin{abstract}
We have carried out a search for radio emission at 820~MHz from six X-ray dim
isolated neutron stars with the Robert C. Byrd Green Bank Radio
Telescope. No transient or pulsed emission was found 
using fast folding, fast
Fourier transform, and single-pulse searches. 
The corresponding flux limits are about
0.01~mJy for pulsed emission, depending on the integration time for
the particular source and assuming a duty cycle of 2\%, and
20~mJy for single dispersed pulses.  These are the most sensitive
limits to date on radio emission from X-ray dim isolated neutron stars.
There is no evidence for isolated radio pulses, as seen in 
a class of neutron stars known as rotating radio transients.
Our results imply that either the radio luminosities of these objects are lower than those
of any known radio pulsars, or they could simply be
long-period nearby radio pulsars with high magnetic fields beaming away from the Earth.
To test the latter possibility, we would need around
40 similar sources to provide a $1\sigma$ probability of at least one
of them beaming toward us. We also give a detailed description of our implementation
of the Fast Folding Algorithm.
\end{abstract}


\section{Introduction}\label{intro}

The discovery of two unusual X-ray sources, RX~J1856.5$-$3754 
and RX~J0720.4$-$3125, by the {\em ROSAT} satellite in the mid-1990s
\citep{walter1996,haberl1996,haberl1997} gave rise to a new class of isolated
radio-quiet neutron stars that are commonly known as X-ray dim
isolated neutron stars (XDINS). To date, a total of seven such
sources are known\footnote{These seven unique objects are sometimes
called the ``Magnificent Seven'' 
as their number has remained
constant since 2001, despite extensive searches. However, the
recently discovered isolated
compact object 1RXS J141256.0+792204, dubbed ``Calvera'',
 may also belong to this group \citep{rutledge2008}.},
all discovered in
the {\em ROSAT} All-Sky Survey \citep{voges1999}, which share very
similar properties \citep[see][]{haberl2004, haberl2007, kaplan2008}.
They are related to the local overabundance of young stars, known
as the Gould Belt \citep{popov2003}.

XDINSs are characterized by soft blackbody-like spectra with
temperatures $kT\!\sim\!40$--$100$~eV, with no indication of
harder, non-thermal components. Their X-ray fluxes are very stable,
with stringent limits of a few percent on long-term variability
\citep{walter1996, haberl1997, motch1999}. 
However, long-term spectral variations have recently been found 
for RX~J0720.4$-$3125 \citep[][~and references therein]{haberl2006}.
Very faint optical
counterparts have been detected for six of the XDINSs to varying confidence levels
\citep{walter1997, motch1998, kulkarni1998, haberl2004a,kaplan2002,kaplan2003b,zane2008,schwope2009}
with extremely large X-ray to optical flux ratios of
$f_\mathrm{x}/f_\mathrm{opt}\gtrsim 10^4$ 
\citep[see, e.g.,][~and references therein]{schwope1999}.
Limits on optical
counterparts for the other known XDINSs correspond to
$f_\mathrm{x}/f_\mathrm{opt}$ of at least $10^3$ \citep[e.g.,][]{zampieri2001}.
Such large values
strongly suggest that XDINSs are isolated neutron stars \citep{maccacaro1988, treves2000}.

Low values of hydrogen column density $N_\mathrm{H} \sim
10^{20}$~cm$^{-2}$ derived from their X-ray spectra signify that
XDINSs are nearby objects not further than 1~kpc away \citep{posselt2007}. This is
confirmed by parallax measurements \citep{vanKerk2007,kaplan2007} 
for two stars, RX~J1856.5$-$3754 and
RX~J0720.4$-$3125, with derived distances of
$161^{+18}_{-14}$ and $360^{+170}_{-90}$~pc. XDINSs
therefore appear to be faint objects with X-ray luminosities
in the range $10^{30}$--$10^{32}$~erg s$^{-1}$.

The most probable origin for the soft X-rays observed from XDINSs is
thermal emission from middle-aged ($\sim 10^5$--$10^6$~yr) cooling
neutron stars \citep[see, e.g.,][and references therein]{haberl2007}. 
This makes these objects important laboratories for testing theories of neutron star cooling,
and therefore for nuclear physics.
It seems unlikely that the soft X-rays are due to Bondi-Hoyle accretion from the interstellar medium
because of the large proper motions measured for the three brightest XDINSs 
\citep{walter2002, zane2006, kaplan2007}.

X-ray pulsations with periods in the range 3--12~s have been detected
in at least six XDINSs.
Assuming that these neutron stars were born with
periods of order milliseconds, the current periods imply strong
magnetic fields $B\sim 10^{13}$~G, 
to provide spin-down.
This scenario has obtained support from independent
measurements of $B$. 
First, for all XDINSs but RX~J1856.5$-$375, broad
absorption features at energies $\sim 100$--700~eV were found in their
X-ray spectra \citep{haberl2003, vanKerk2004, haberl2004a,
haberl2004b,zane2005}.  If due to proton cyclotron resonance, 
and/or bound-free or bound-bound transitions in H, H-like, and He-like atoms,
they imply magnetic field strengths of the order of $\sim 10^{13}-10^{14}$~G.
Independent measurements of $1.5\times 10^{13}$, $2.4\times 10^{13}$,
$3.4\times 10^{13}$, and $2\times 10^{13}$~G 
were obtained by \citet{vanKerk_kaplan2008}, \citet{cropper2004}, and \citet{kaplan2005a,
kaplan2005b,kaplan2009} for RX~J1856.5$-$375, RX~J0720.4$-$3125, RX~J1308.6+2127, and RX J2143.0+0654
from the period derivative (determined through phase-connected
timing techniques) under the assumption of magneto-dipolar losses.

Neither associations with supernova remnants nor confident detections
of radio emission have been found for any XDINS thus far.  For
RX~J1856.5$-$3754 a limit of 4~mJy on the pulsed flux from a 430-MHz
observation with the 64-m Parkes radio telescope and a $5\sigma$ upper
limit of 0.6~mJy on the continuum flux at 5~GHz from a 330-s VLA
snapshot observation were reported by \citet{walter1996} 
\citep[see also][]{brazier1999, perlman1996}.
Assuming 3\% duty
cycles for RX~J0720.4$-$3125 and RX~J0806.4$-$4132, $3\sigma$ upper limits
of 0.05~mJy on pulsed radio emission were obtained at 1384 and
1704~MHz using the Australia Telescope Compact Array radio
interferometer \citep{johnston2003}.  A slightly better limit of
0.02~mJy at the $8\sigma$-level (for a duty cycle of 1\%) for the
pulsed flux of RX~J0720.4$-$3125 was estimated by \citet{kaplan2003a}
from Parkes observations at 1374~MHz.  They have also reported an
upper limit at 644~MHz of about 0.2~mJy given the same
detection limit and duty cycle. 
No radio pulsations or isolated pulses were found in a Parkes search
of  RX~J2143.0+0654
at 0.78 and 2.9~GHz reported recently
by \citet{rea2007}. For a pulse duty cycle of 5\%, they obtained flux density 
upper
limits of 0.33 and 0.06~mJy at 0.78 and 2.9~GHz,
respectively.
Recently, \citet{malofeev2005,malofeev2007} reported the
detection of weak radio emission from two XDINSs, RX~J1308.6+2127 and
RX~J2143.0+0654, at the very low frequency of 111~MHz with the
Large Phased Array (BSA) at Puschino Radio Observatory. They measured
flux densities of $50\pm 20$ and $60\pm 25$~mJy, for
RX~J1308.6+2127 and RX~J2143.0+0654, respectively. Though
very intriguing, independent observations with other telescopes at similar 
frequencies are essential to confirm these detections.

\begin{figure}[hbtp]
\includegraphics[scale=0.44]{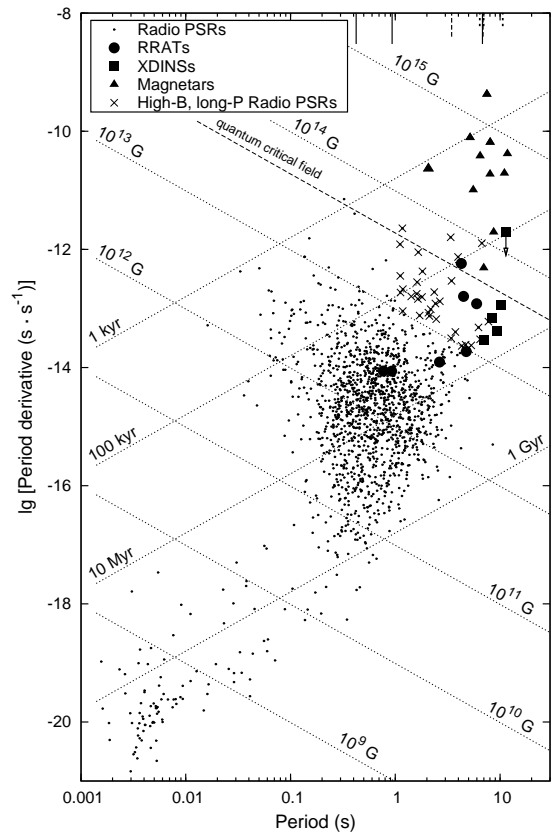}
\caption{\ppdot diagram. Lines on the top mark the period values for
objects with no measured period derivatives: three RRATs (solid),
one soft $\gamma$-repeater, SGR~1627$-$41, and two anomalous X-ray pulsars, AX~J1845$-$03 and
CXO~J164710.2$-$455216 (dashed), and two XDINSs RX~J0420.0$-$5022 and
RX~J1605.3+3249 (longer dashed). 
High-$B$ long-period radio pulsars are those with periods $> 1$~s and
magnetic fields $B >10^{13}$~G (see Section~\ref{discussion}).
Dotted lines are those
for constant inferred dipolar magnetic field strength and constant
characteristic age. The dashed line depicts the quantum critical field
of $4.4\times 10^{13}$~G.
}
\label{fig1}
\end{figure}

There are many striking similarities between X-ray dim isolated
neutron stars, magnetars, rotating radio transients (RRATs)
\citep{mmclaugh2006}, and high-$B$ long-period radio pulsars.
They occupy similar, but not identical, overlapped regions 
in the period ($P$) and period derivative ($\dot{P}$)
diagram shown in Fig.~\ref{fig1}, with 
ages and magnetic fields suggesting evolutionary
relationships. 

Most of the RRATs were not yet observable at X-ray energies
due to their poor position localization. At present, however, seven out
of eleven RRATs have timing solutions and hence positional
accuracies suitable for X-ray observations \citep{mmclaugh2009}.
X-ray pulsations were already detected from the 4.26-s RRAT~J1819--1458
in a 43-ksec XMM-{\em Newton} observation
\citep{mmclaugh2007}.  The X-ray pulsations and inferred blackbody
temperature are consistent with the properties of both the XDINSs and
of normal X-ray detected radio pulsars. However, the spectral feature
detected is similar to those seen in the spectra of XDINSs
\citep{vanKerk2007}. The X-ray properties of this source are
also similar to those of the transient magnetar XTE~J1810--197
in quiescence \citep{ims+04,gvbb04}.
However, the radio emission characteristics of these two neutron stars
appear to be quite different.

\placefigure{fig1}

An appealing option, then, is that RRATs are simply more distant
XDINSs \citep{popov2006}. Being that RRATs are at much larger distances (a few kpc, as
derived from their dispersion measures (DMs)
compared to a few hundred pc for the XDINSs), their dim X-ray emission would not
have been detectable (and, indeed, was not detectable for J1819--1458) in
 the ROSAT all-sky survey. If this hypothesis is
correct, the putative transient radio emission from XDINSs should be
easily observed. However, to our knowledge, all previous
radio investigations, except the recent observation of RX~J2143.0+0654 
\citep{rea2007}, only focused on search for periodic radio pulses, like 
those seen in ordinary pulsars. In this case, any
RRAT-like radio emission could have escaped detection.

In this paper we attempt to directly test this hypothesis, and report
on the results of an 820-MHz search for periodic and transient radio
emission from six out of the seven known XDINSs visible from with the Robert
C.~Byrd Green Bank Telescope (GBT). We searched for periodic emission
using the standard Fast Fourier Transform (FFT) algorithm, and also included
a Fast Folding Algorithm (FFA), known to be more sensitive to long-period
signals. We also incorporated a search for single dispersed pulses.

We describe the observations in Section~\ref{obs} and data reduction
techniques in Section~\ref{reduction}.  The results of our searches
are presented in Section~\ref{results}. 
The implications of these results
are discussed in Section~\ref{discussion}. Finally, in the Appendix,
we give a brief review of the FFA used to search for pulsed radio emission, 
and provide a
freely-available program which implements the search.

\section{Observations}\label{obs}

Observations of six XDINSs at a center frequency of 820~MHz were
carried out with the GBT from 26 May 2006 to 2 June 2006. Two
orthogonal polarizations were summed and the total intensity was
recorded using the pulsar SPIGOT backend \citep{spigotref}, a
digital correlator which records 1024 3-level autocorrelation functions across a
50~MHz band, sampled every $81.92~\mu$s.  The system temperature of the
0.68--0.92~GHz prime focus receiver is 25~K. We observed all six
XDINSs that are visible at the GBT.  Several test pulsars were also
observed for short exposures for setup and calibration, and to
check the search procedures that we applied to the XDINSs. Among those
observed was the 1.24-s pulsar J0628+09, discovered through a
search for single pulses in the
Arecibo PALFA survey \citep{cordes_palfa2006}.  The list of sources is
given in Table~\ref{tab1} with the modified Julian day (MJD)
of observation and the total observing time.

The choice of 820~MHz was determined by the following considerations.
The spectral indices of XDINSs are unknown. However, their
non-detection so far at radio frequencies $>1$~GHz together with the
possible recent detection of pulsed radio emission from two XDINSs at
111~MHz \citep{malofeev2005, malofeev2007} suggest that the radio
spectra are steep (see Section~\ref{discussion}),
making lower frequencies preferable. In addition,
frequencies around 1400~MHz at the GBT are more affected by radio
frequency interference (RFI). The ideal choice of 350~MHz, 
which  is less affected by RFI, was impossible due to receiver interchange
issues at the time of the observations.

Each XDINS was observed two or three times, with each observation
typically lasting for 4 hours.  However, two out of three
observing sessions of RX~J0806.4$-$4123 were only 2.5 and 2 hours
long, and one of three observing sessions of RX~J1308.6+2127 was only
2 hours long. The optimal duration of 4 hours was chosen as to be
greater than the maximum average interval between consecutive pulses
for the RRAT sources. This is about 3 hours for RRAT~J1911+00
\citep{mmclaugh2006}.

\placetable{tab1}
\begin{table}[tbh]
\caption{Observations summary. \label{tab1}
}
\begin{center}
\begin{tabular}{lcccc}
\hline
\hline
Source & MJD & $\mathrm{T}_\mathrm{obs}$ & $N_\mathrm{s}$ & $T_\mathrm{sky}$ \\
       &     & (h)                       &                & (K) \\
\hline
\\[0.5pt]
\multicolumn{5}{c}{XDINSs} \\
\\[0.5pt]
RX J0720.4$-$3125 & 53881 & 3.84 & x & 7 \\
           & 53883 & 4 & 1 & \\
RX J0806.4$-$4123 & 53882 & 4 & x & 9 \\
           & 53884 & 2.5 & 1 & \\
           & 53885 & 2 & 1 & \\
RX~J1308.6+2127\tablenotemark{a} & 53881 & 4 & x & 6 \\
         & 53883 & 4.25 & 2 & \\
         & 53885 & 2 & x & \\
RX J1605.3+3249\tablenotemark{b} & 53882 & 4 & x & 7 \\
         & 53884 & 4 & 1 & \\
RX J1856.5$-$3754 & 53886 & 4 & 3 & 11 \\
           & 53888 & 4 & x & \\
RX J2143.0+0654\tablenotemark{c} & 53886 & 4.08 & 3 & 7 \\
         & 53888 & 4 & x & \\
\\[0.5pt]
\multicolumn{5}{c}{Radio Pulsars} \\
\\[0.5pt]
B0540+23 & 53881 & 1~min & 1 & 11 \\
B1534+12 & 53884 & 15~min & 1 & 10 \\
  & 53886 & 2~min & 1 & \\
J0628+09 & 53885 & 2 & 1 & 9 \\
\hline
\end{tabular}
\end{center}
\tablecomments{$T_\mathrm{obs}$ is the total duration of the observing
run in hours, $N_\mathrm{s}$ the number of sections into which the
data were split due to jumps in baseline. The duration of each section
is given as $t_\mathrm{run}$ in Table~\ref{tab2}. The symbol 'x'
means that these data were excluded from our analysis due to large
contamination by RFI. $T_\mathrm{sky}$ is the sky temperature
extrapolated from the all-sky 408-MHz survey by \citet{haslam1982}
using a spectral index for the sky background radiation of $-2.6$.
The contribution 
of the cosmic microwave background is also taken into account.}  
\tablenotetext{a}{~RBS~1223}
\tablenotetext{b}{~RBS~1556}
\tablenotetext{c}{~RBS~1774}
\end{table}

\section{Data Reduction}\label{reduction}

All the processing described here was carried out on a Beowulf cluster
at West Virginia University. The raw autocorrelation functions for
each time sample recorded by the SPIGOT were first corrected for
3-level quantization biases before being Fourier transformed to
synthesize 1024 frequency channels across the 50~MHz band. To reduce
the computational requirements in our search, where predominantly
long-period signals are expected, the data were subsequently
downsampled by a factor of six for an effective time resolution of
$491.52~\mu$s using the
\emph{SIGPROC}\footnote{\url{http://sigproc.sourceforge.net}}
package. To characterize the RFI environment during our observations, we
used the \emph{rfifind} program from the
\emph{PRESTO}\footnote{\url{http://www.cv.nrao.edu/~sransom/presto}}
package on all data. The program searches for strong broad-band
outbursts and periodic interference and creates a mask which can be
applied to further processing.

\placetable{tab2}
\begin{table*}[btp]
\caption{XDINSs processing summary\label{tab2}}
\begin{center}
\begin{tabular}{ccccccccc}
\hline

\hline
XDINS & $P$ & $\dot{P}$ & d & $t_\mathrm{run}$  & $N^\mathrm{ch}_\mathrm{ignored}$ & DM$_\mathrm{max}$ & DM$_\mathrm{step}$ & Ref \\
       & (s) & ($10^{-12}$ s s$^{-1}$) & (kpc)  & (s) & (\%) & (pc cm$^{-3}$) & (pc cm$^{-3}$)  & \\
\hline
RX J0720.4$-$3125 & 8.39  & 0.0698 & 0.36  & 14400 & 13   & 98   & 1.21 & 1,2 \\
RX J0806.4$-$4123 & 11.37 & $<2$   &       & 9000  & ---  & 172  & 0.7  & 3,4\\
                  &       &        &       & 2321  & 20   &      & 1.45 & \\
RX~J1308.6+2127  & 10.31 & 0.112   &       & 7923  & 0.1  & 27   & 1.21 & 5,2 \\
                  &       &        &       & 7297  & 0.1  &      &      & \\
RX J1605.3+3249   & 6.88?  & ---    &       & 14400 & ---  & 31   & 0.6  & 6\\
RX J1856.5$-$3754 & 7.055 & 0.0297 & 0.161 & 1680  & ---  & 37   & 0.6  & 7,8 \\
                  &       &        &       & 6002  &      &      &      & \\
		  &       &        &       & 3741  &      &      &      & \\
RX J2143.0+0654   & 9.437 & 0.04   &       & 2961  & 12   & 33   & 1.21 & 9,10 \\
                  &       &        &       & 1441  & 12   &      &      & \\
		  &       &        &       & 5682  & 15.4 &      &      & \\
\hline
\end{tabular}
\tablecomments{Column 4 lists the parallax distances for two XDINSs
from \citet{vanKerk2007} and \citet{kaplan2007}.  However, for all
neutron stars the values of DM$_\mathrm{max}$ were obtained assuming
distances of 1~kpc and then doubled. For radio luminosity
estimation in Table~\ref{tab3} we used distances listed
in column 4 for two sources and a conservative value of
1~kpc for the other sources. The values of $P$ and $\dot{P}$ are from the papers
given in  column 9.}
\tablerefs{(1) \citet{haberl1997} (2) \citet{kaplan2005a} (3) \citet{haberl2002} (4) \citet{haberl2004a} 
(5) \citet{haberl2003} (6) \citet{haberl2007} (7) \citet{tiengo2007} (8) \citet{vanKerk_kaplan2008} 
(9) \citet{zane2005} (10) \citet{kaplan2009} }
\end{center}
\end{table*}

\begin{figure}[hbt]
\includegraphics[angle=270,width=\columnwidth]{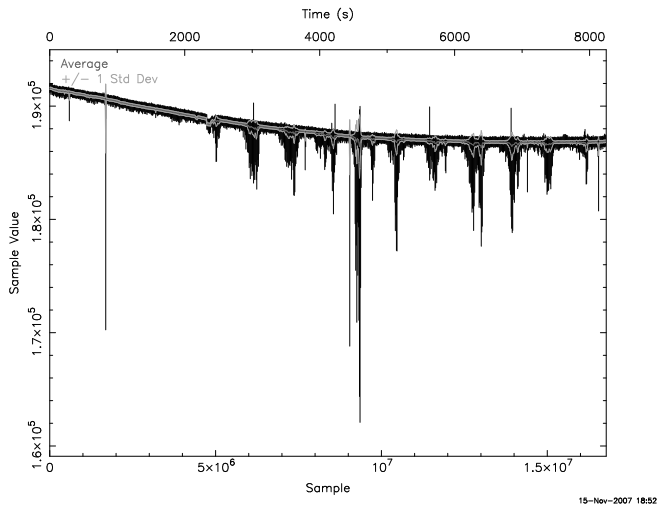}
\caption{An observation of
RX~J0806.4$-$4123 severely
affected by RFI. Shown is the flux density in arbitrary units versus
the time (top axis) or sample number (bottom axis) for about 8000~s of
a 4-h observation. The plot was made by dedispersing the 
data with DM$=0$~pc cm$^{-3}$ using the \emph{prepdata} program from the
\emph{PRESTO} package. Gray lines represent the average and $\pm
1\sigma$ contours. Each of these quasi-periodic negative bursts
consist of many shorter leaps.
}
\label{fig2}
\end{figure}

Almost all of the data were severely affected by RFI, with half of it
(see Table~\ref{tab1}) of such bad quality that we excluded them
from our analysis.  Fig.~\ref{fig2} shows an example of one
of the omitted observations. All RFI seemed to be of either equipment
and external nature.\footnote{See the GBT PF1 800 MHz RFI Survey at
\url{http://www.gb.nrao.edu/~tminter/rfi/PF800-rfi.shtml} for the list
of known RFI in this frequency band.} In addition to the large amount
of broad-band impulsive RFI, we also measured a large feature at
around 818~MHz having a bandwidth of about 2.5~MHz. This is a resonant
frequency from the orthomode transducers that split the
polarizations.\footnote{\url{http://www.gb.nrao.edu/electronics/GBTelectronics/Receivers/prime.html}}
This strong feature affects about 100 channels in almost all the
data so we excluded them from our processing.

\placefigure{fig2}

\begin{figure*}[hbt]
\includegraphics[angle=270,width=\columnwidth]{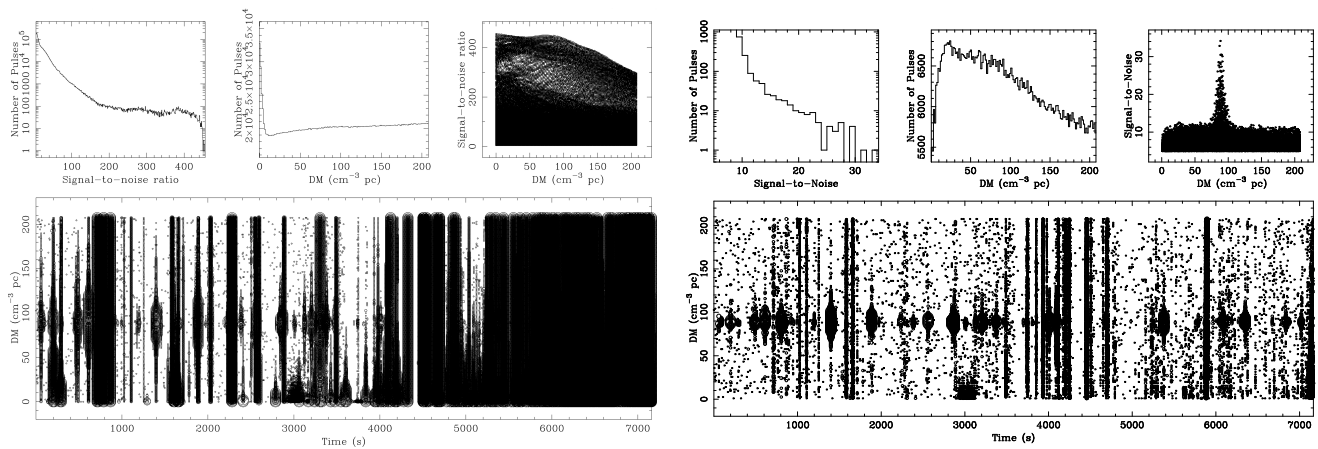}\vskip -59mm\hskip 90mm\includegraphics[angle=270,width=\columnwidth]{f3b.eps}
\vskip 1mm
\caption{
The single-pulse search diagnostic plots for the 1.24-s pulsar
J0628+09 without RFI excision (left) and with the manual RFI excision
technique applied (using \emph{rfimarker}, right). They have four
smaller insets.
{\it Top:} the histograms of the number of pulses
versus S/N and DM, and the dependence of S/N versus DM {\it (from left
to right)}.  {\it Bottom:} the main time-DM plot with the size of the points
corresponding to the S/N of the pulse.  The detection threshold is
$5\sigma$. Vertical lines of pulses are either stronger at zero DM or
roughly constant over all the DM range. They are RFI signatures rather
than a real pulse, which would appear as a vertical segment of pulses
with S/N peaking at a non-zero DM and
gradually fading towards the edges of the segment (see bottom plot of the Fig.~\ref{fig4} how the pulse signatures appear
for the low-DM fake pulsar).
The train of pulses
is clearly seen in the right plot (bottom inset). The strong peak in
S/N-DM inset confirms the detection.
}
\label{fig3}
\end{figure*}

\subsection{RFI Excision}\label{rfiexcision}

For the half of the data that were not excluded, we created a mask to
zap the RFI by using the \emph{rfifind} package from
\emph{PRESTO}. Zapped samples were substituted by the mean of 80\% of
samples (excluding outliers) in every 30~s long chunk of data in
each channel.  However, even after applying the mask, the time series
at dispersion measure 
$\mathrm{DM}=0$~pc cm$^{-3}$ and at larger values exhibited large outliers,
fluctuations and jumps in the baseline levels. This is because
\emph{rfifind} deals only with strong broad-band outbursts by
clipping them at $\mathrm{DM}=0$~pc~cm$^{-3}$ (all spikes stronger than
$6\sigma$ by default) and with periodic RFI (Scott Ransom, private
communication).  However, there were still many interference
signatures left in the data of mostly non-periodic nature.  In
addition, some data were influenced by jumps in the baseline that
could not be corrected by \emph{rfifind}.

First, to overcome the jumps in power levels we divided the
corresponding data into several sections and processed them
separately. The fourth column in Table~\ref{tab1} shows the
number of sections that were used for every observation.  The duration
of each observation is given in Table~\ref{tab2} (column 5).
Then, we dedispersed every piece with $\mathrm{DM}=0$~pc cm$^{-3}$ and
inspected these time series for RFI. To do this the
\emph{rfimarker} program was written.  All records
were inspected manually, and RFI-affected samples were selected by eye and
flagged using a cursor.  
Then another program \emph{rficut} read both
the input data file and RFI binary list produced by
\emph{rfimarker} and replaced the bad samples by the average in a
window of 131072 samples, excluding the top 10\% of outliers.
We also
inspected the bandpass and if there were some frequency
channels strongly affected by RFI, they were excluded from
further analysis. This was done by providing a list of channels to ignore as an
input option to the dedisperse program from the \emph{SIGPROC} package. The
percentage of channels that were ignored is listed in Table~\ref{tab2}
(column 6).  The advantage of RFI excision is shown in
Fig.~\ref{fig3} where the diagnostic plot for the single-pulse
search for pulsar J0628+09 is shown both for original data and the
data with the masked RFI processed with \emph{rfimarker}.

\begin{figure}[hbt]
\includegraphics[angle=270,width=\columnwidth]{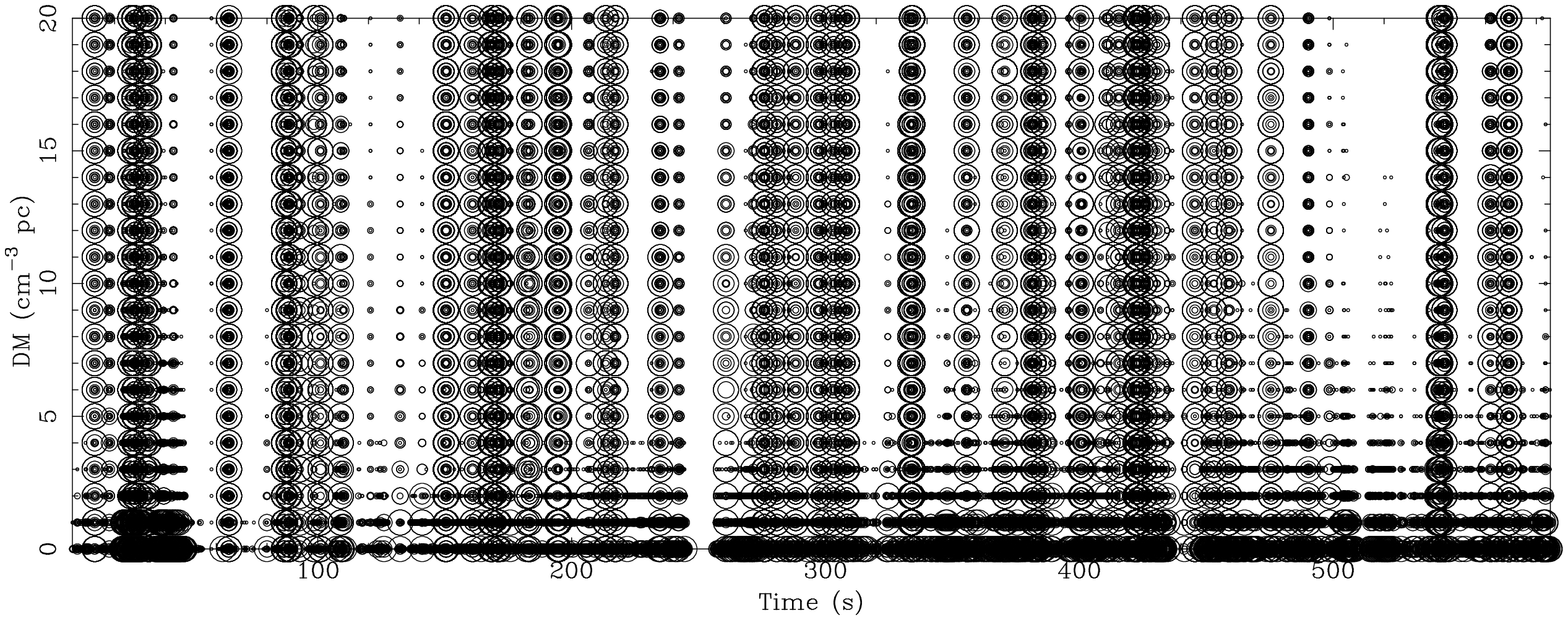}
\includegraphics[angle=270,width=\columnwidth]{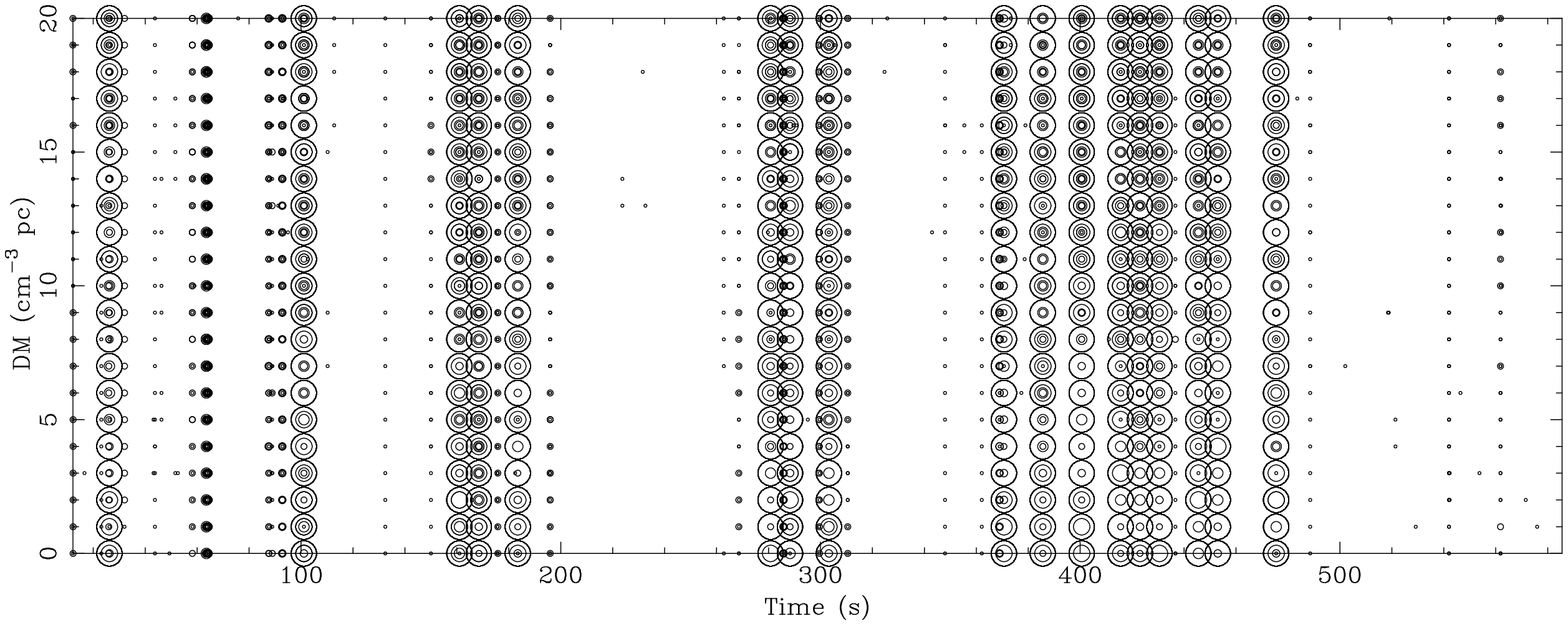}
\caption{The time-DM plots from the single-pulse search analysis for the fake pulsar signal
injected into real data without RFI excision (top) and with the RFI clipping technique applied (bottom).
The RFI excision technique is described in Section~\ref{rfiexcision}. These plots are similar
to the bottom insets of Fig.~\ref{fig3}. Three other similar insets do not provide any necessary
information and are not presented here. The real data used to inject the fake pulsar signal is the
chunk of length 600~s from the observation of pulsar J0628+09 offset
by 6000~s from the start of the observation (see Fig.~\ref{fig3}). This particular chunk was greatly
affected by RFI. The range of DMs processed
is 0--20~pc cm$^{-3}$, thus the contribution from the pulsar J0628+09 with DM$=88$~pc cm$^{-3}$ is negligible.
The parameters of the injected fake pulsar are: $P = 7510.686$~ms, DM$=3.27$~pc cm$^{-3}$, duty cycle $= 0.315$\%,
peak S/N $= 10.3$, and null fraction = 79\%. The vertical lines of large circles are true fake pulses
rather than RFI.
}
\label{fig4}
\end{figure}

A careful reader may notice that in our paper we are focusing 
on searching for low-DM signals expected from nearby XDINSs rather than
higher DM signals such as pulsar J0628+09. In the case of a low-DM source, the
chance of occasionally zapping a real pulse from an XDINS is higher. Furthermore,
in plots similar to Fig.~\ref{fig3},
it would be more difficult to distinguish real pulses from zero-DM RFI.
This is why the visual inspection aspect of {\it rfimarker} is important; 
much of the RFI has repeatable characteristic signatures that
the  trained eye will recognize. To show that
\emph{rfimarker} does not zap real pulses, we  injected a fake pulsar signal into the
real data shown in Fig.~\ref{fig3} contaminated by RFI. 
We cleaned these data with \emph{rfimarker}, processed it
in a range of DMs and made a diagnostic single-pulse plot. 
We chose a subset of data severely affected by RFI, namely
a series of 600~s duration starting roughly 6000~s from the beginning of the observation
(see Fig.~\ref{fig3}). The range of DMs processed was 
0--20~pc cm$^{-3}$; therefore the contribution from the pulsar J0628+09 itself, with 
DM$=88$~pc cm$^{-3}$, is negligible.
The parameters of the fake pulsar were randomly
generated from the range of the values we might expect for XDINSs, namely
$P$ from 5--10~s, DM from 3--10~pc cm$^{-3}$, duty cycle from 0.1--2\%, null fraction from
50--90\% and peak SNR from 7--15.
The results are shown in Figure~\ref{fig4}.
Without RFI excision,
one can see many RFI peaks across the entire range of  DMs with a larger contribution
at lower DMs as expected. With RFI zapping,
the plot looks cleaner, but one still
can see several vertical lines similar to those we expected from RFI. 
They are, however, actually the pulses of the injected fake pulsar. After the injection of the fake
pulsar and clipping we checked the parameters of the injected pulsar and how many
pulses were injected (we did not know them in advance). 
There were 17 of them, in agreement with the null fraction for this fake pulsar,
its period and data length. The number of these vertical lines is the same.
The time differences between them all are equal to the integer number of periods of the fake pulsar
(we used this technique for real XDINSs data, see Section~\ref{spsearch}).
We did this experiment for several fake injected pulsars, with similar results. 
Therefore, this confirms that our manual RFI clipping technique 
does not zap real pulses. 

However, as was mentioned above, it is difficult to 
distinguish between real low-DM pulses and RFI on the single-pulse diagnostic plots.
This is because it is hard to tell whether the S/N peaks at a low DM or zero DM.
Using Eq. 12 from \citet{cordes_mam2003}, the ratio
of the measured S/N of the pulse, when dedispersed with the wrong DM, to the true S/N,
is almost 1 
for a pulse width of 10~ms and error in  DM of $1$~pc cm$^{-3}$ (given the observing
frequency and total bandwidth of our observations). 
For a larger error in DM of $10$~pc cm$^{-3}$, this ratio is 
only 0.88 for a 10-ms pulse, and closer to 1 for broader pulses which we might expect from XDINSs.
In any case, the broadening due to dispersion
conserves the pulse area, and due to the matched filtering techniques we would get almost the same S/N,
but with a larger effective sampling time, equal to the observed width of the pulse. Thus, there is no
significant decrease in S/N for low-DM signals when one inspects the single-pulse diagnostic plots.
All of the above considerations are true for high-DM pulsars as well,
but in the diagnostic plots we inspect a much
larger range of DM for them where the decrease in S/N at zero DM is more noticeable.

To combat this difficulty, we used the same technique which was used
by \citet{mmclaugh2006} and resulted in the discovery of the rotating radio transients.
The results of this search is given below in the Section~\ref{spsearch}.

\placefigure{fig3}

\subsection{Search pipeline}

After the excision of RFI, each corrected observation was dedispersed
for a number of different dispersion measures, DMs, from zero to a
maximum value DM$_\mathrm{max}$ (column 7, Table~\ref{tab2}) with a
step size of roughly 1~pc~cm${}^{-3}$ (column 8). The DM spacing was
determined by considering the dispersion smearing of $491.52~\mu$s (1
sample) in one frequency channel and the minimum pulse width of
1~ms.\footnote{Pulsars are known to have a broad inverse correlation between
the duty cycle and their period~\citep[see, e.g.,][]{lyne1988}. The smallest
known duty cycle of $\approx 0.03\%$ from the ATNF catalog belongs to the 6.8-s
RRAT J1848$-$12. If an XDINS from this work would have the same duty cycle, this would imply
the pulse width of about 2~ms for the RX J1605.3+3249, with the shortest period
in our list. Thus, we chose the conservative value of 1~ms in our determination
of DM spacing.}
We chose the maximum value DM$_\mathrm{max}$ as double the prediction
from the NE2001 model of free electron density in our
Galaxy~\citep{ne2001} assuming distances up to 1~kpc for all sources.
The corresponding searched distance ranges are
certainly much larger than any distance estimates for XDINSs \citep{posselt2007} or known
parallax distances reported by \citet{vanKerk2007} and
\citet{kaplan2007}.  Once the data were dedispersed they were then
searched for single pulses using the single-pulse search
technique described in \citet{cordes_mam2003}.  The data were searched for
periodic emission as well using two search techniques, the FFT and the FFA.
Both 
single-pulse and FFT searches are incorporated in the \emph{SIGPROC} pulsar
software package.
We wrote an FFA search code \emph{ffasearch}
to implement the FFA algorithm (see 
Appendix).  This code has been
proven to detect known long-period pulsars with higher signal-to-noise
ratios (S/N) than traditional FFT searches.

\begin{table*}[btp]
\caption{XDINSs Results\label{tab3}}
\begin{center}
\begin{tabular}{ccccccc}
\hline

\hline
 & \multicolumn{3}{c}{Pulsed emission}   & \multicolumn{3}{c}{Bursty emission} \\
\cline{2-4}\cline{5-7}
XDINS & $S_\mathrm{lim}$ & $L_{1400}^\mathrm{p,max}$ & $L_{820}^\mathrm{p,max}$ & rate upper limit & $S_\mathrm{lim}^\mathrm{sp}$ & $L_{1400}^\mathrm{b,max}$\\
      & ($\mu$Jy) & (mJy kpc$^2$) & (mJy kpc$^2$) & (hr$^{-1}$) & (mJy) & (mJy kpc$^2$) \\
\hline
RX J0720.4$-$3125 & $8$   & $4\times 10^{-4}$ & $10^{-3}$ & $0.25$ & $21$ & $1$    \\
RX J0806.4$-$4123 & $10$  & $4\times 10^{-3}$ & $10^{-2}$ & $0.32$ & $18$ & $6.9$ \\
RX~J1308.6+2127   & $10$  & $4\times 10^{-3}$ & $10^{-2}$ & $0.24$ & $17$ & $6.5$ \\
RX J1605.3+3249   & $8$   & $3\times 10^{-3}$ & $8\times 10^{-3}$ & $0.25$ & $22$ & $8.4$ \\
RX J1856.5$-$3754 & $14$  & $1.4\times 10^{-4}$ & $3.6\times 10^{-4}$ & $0.32$  & $24$ & $0.2$   \\
RX J2143.0+0654 & $13$ & $5\times 10^{-3}$ & $1.3\times 10^{-2}$ & $0.36$ & $20$ & $7.6$   \\
\hline
\end{tabular}
\end{center}
\vskip 3mm
\end{table*}

\placefigure{fig4}

\section{Results}\label{results}

No pulsed radio emission or isolated pulses were found from any of the six observed
isolated neutron stars.  Table~\ref{tab3}
summarizes the results. Column 1 lists the name of the
source, column 2 gives the $4\sigma$ flux upper limits
$S_\mathrm{lim}$ on pulsed radio emission for an assumed pulse duty
cycle of 2\%, and column 3 lists the upper limit on radio luminosity
at 1400~MHz, $L_{1400}^\mathrm{p,max}$, in mJy kpc$^2$ assuming 
spectral index\footnote{The spectral index
$\alpha$ is defined here as $S_\nu \propto \nu^{\alpha}$, where
$S_\nu$ is the flux density at frequency $\nu$.} $\alpha = -1.8$
and either measured parallax distances
from Table~\ref{tab2} (column 4) for two XDINSs or a very conservative
distance estimate of 1~kpc.  Column 4 gives the radio
luminosity at our observing frequency of 820~MHz,
$L_{820}^\mathrm{p,max}$, with the same distance assumptions as
for column 3.  Columns 5--7 summarize the results on single-pulse
emission and list the upper limits on single-pulse detection rate, peak fluxes
$S_\mathrm{lim}^\mathrm{sp}$ and radio luminosities
$L_{1400}^\mathrm{b,max}$ with the same assumptions as for radio
luminosity for pulsed emission. The spectral indices of XDINSs are
unknown, and for our estimates we used the
mean value of $-1.8$ measured for known radio pulsars \citep{maron2000}.

\placetable{tab3}

\subsection{FFT Search}

Though the FFT search is not efficient for long period sources (see~\ref{simul_section}),
we still applied it to every observed
XDINS. We tried three incoherent harmonic summations with 32, 64, and
128 harmonics. However, we did not find any promising candidates.

\subsection{FFA search}

\begin{figure}[hbt]
\includegraphics[angle=270,width=\columnwidth]{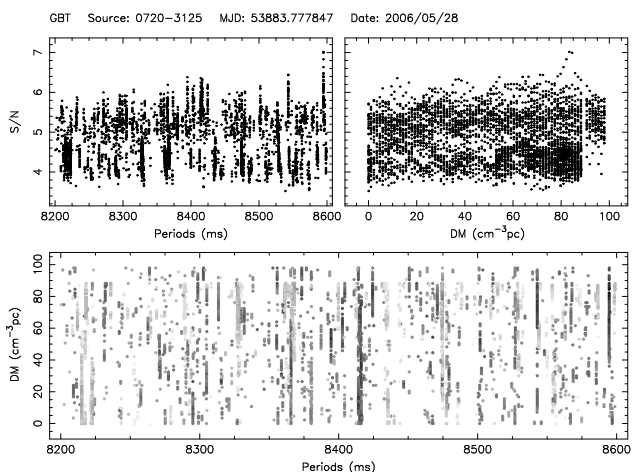}
\caption{Example of an FFA diagnostic plot for RX~J0720.4$-$3125.  The
{\it top-left} plot is the periodogram, i.e.~the folded profile's
significance for every periodic candidate, the {\it top-right} plot
shows the profile significance versus DM. The plot on the {\it bottom}
represents the S/N of the folded profiles versus DM and the
period. The darker points have larger S/Ns. The expected signature of
the pulsar is a bar of increased flux along the DM axis (see
Appendix
for more details). Long lines over a broad DM range are not real but
rather due to strong RFI of periodic origin.
}
\vskip 3mm
\label{fig5}
\end{figure}

On the contrary, the FFA search should be very effective for long-period pulsars. The
\emph{ffasearch} program was written to implement the FFA algorithm
(see 
Appendix). We performed this search for period ranges within
200~ms of the reported X-ray period (see Table~\ref{tab2}).  These
ranges are much larger than the difference between the reported
barycentric X-ray period value and the topocentric period at the date
of observation even for a period derivative as large as  $\dot{P} = 10^{-11}$~s s$^{-1}$.
They are also much larger
than any reasonable
uncertainties in the X-ray periods.  Following the matched filtering
technique by \citet{cordes_mam2003}, every profile was rebinned by
different factors from 1 (no rebinning; sampling interval is
$491.52~\mu$s) to $N$, where $N$ corresponded to a sampling interval
approximately equal to 0.04$P$, where $P$ is the X-ray period. Thus,
our search was sensitive for all possible pulse widths up to duty
cycles of 4\%.  We have inspected the folded profiles for many
candidates and did not find any significant profiles down to a
$4\sigma$ detection threshold. 
Those profiles with peak fluxes still above a $4\sigma$ level
were undoubtedly of RFI nature because they
 were detected over a broad range of DMs including $\mathrm{DM} = 0$ pc
cm$^{-3}$. We therefore quote a $4\sigma$ upper limit on pulsed radio
emission.  As an example, Fig.~\ref{fig5} shows an FFA diagnostic
plot for RX~J0720.4$-$3125. The presence of a real pulsed source would
show up as an elongated train of candidates in both period and DM axes.
\vskip 3mm

\placefigure{fig5}

\subsection{Single-Pulse Search}\label{spsearch}

The single-pulse search is a very powerful tool to detect
strong individual pulses from sources whose regular, periodic emission is too weak
to be detected through a periodicity search. The RRATs were discovered
in this way~\citep{mmclaugh2006}. XDINSs show possible connections with
the RRATs, thus making the single-pulse search very important.

We performed a single-pulse
search for each isolated neutron star in our analysis.
The search was done for a range of pulse widths using the matched
filtering technique~\citep{cordes_mam2003} up to maximum widths of about
503~ms (ten consecutive summations of neighboring samples).
If pulses were detected with several different matched filters, only the
candidate with the largest S/N was recorded. For every XDINS we obtained 
the candidates list of all events stronger than $5\sigma$ and single-pulse
diagnostic plots\footnote{The 
diagnostic plots for the single-pulse search can be found at \url{http://astro.wvu.edu/projects/xdins}.}
similar to those in Fig.~\ref{fig3},~\ref{fig4}.
However, as was discussed in the Section~\ref{rfiexcision} it is very hard to distinguish
real low-DM signals from zero-DM RFI in these plots.

To overcome this difficulty, we can make use of the fact that any pulses from an XDINS should be
emitted modulo the known rotation period (see Table~\ref{tab2}). 
By searching for the greatest common denominator
of the differences between all of the pulses, we can determine whether they have an underlying
periodicity. This was the method used to reveal the periods of the rotating radio transients \citep{mmclaugh2006}. Due to the inevitable presence of some pulses due to RFI, we simply report the period that fits
the greatest number of differences, but do not require that all pulses fit. We applied this method to the test pulsar
J0628+09 and the fake pulsar from Fig.~\ref{fig4}. For the pulsar J0628+09 we found that, despite the presence of pulses due to RFI, the underlying pulsar periodicity was evident. The fake pulsar periodicity was found in the 
clipped data only, emphasizing the importance of the RFI mitigation. Applying this method to
the single pulses detected in this XDINS search revealed no obvious periodicities. Because we know the
periods of the XDINSs a priori, we also performed another search in which we took all of the differences between pulses and divided by the known period. There was no evidence that any of the differences between detected pulses were
integer multiples of the spin period. 

We therefore can conclude with confidence that our single pulse search did not detect 
any pulses from the XDINSs. 

\subsection{Flux Upper Limits}

The minimum detectable flux of our survey strongly depends on the
pulse duty cycle, $\delta$, as
\begin{equation}
S_\mathrm{lim} = 354\frac{T_\mathrm{sys}\cdot\mathrm{S/N}}{\sqrt{t_\mathrm{run}~\!B}}\sqrt{\frac{\delta}{1-\delta}}~\mu \mathrm{Jy}
\label{eq1}
\end{equation}
for our observing setup at 820~MHz. In this expression,
$T_\mathrm{sys}$ is the system temperature in K including sky
contribution, $t_\mathrm{run}$ the observing time in seconds, $B$ the
effective bandwidth in MHz, and S/N the detection threshold.
Assuming a reasonable value of $\delta = 2\%$, greater than the duty cycle of 65\% of 
pulsars with $P > 2$~s, our $4\sigma$ detection limits
range from 8 to $14~\mu$Jy for different XDINSs
depending on the length of the observation and number of ignored
frequency channels, i.e. the effective bandwidth.  The list of the
flux upper limits is given in Table~\ref{tab3} (column~2).
The single-pulse duty cycles of the RRATs are even smaller than 2\%, with the mean value around
0.5\% \citep{mmclaugh2006, mmclaugh_iau}.  If the integrated duty cycles of
possible radio emission from XDINSs are the same as for RRATs, then the
flux limits are even smaller than those listed.

For single radio pulses, the flux limits can be obtained from
\begin{equation}
S_\mathrm{lim}^\mathrm{sp} = 11.2\frac{T_\mathrm{sys}\cdot\mathrm{S/N}}{\sqrt{w~\!B}}~\mathrm{mJy}~,
\end{equation}
where $w$ is the assumed pulsed width, and other quantities are the
same as in equation~(\ref{eq1}).  Assuming pulse widths $w = 0.02P$,
where $P$ is the X-ray period, we derive $5\sigma$ upper
limits on single-pulse fluxes ranging from 17 to 24~mJy. This list is given
in column~6 of Table~\ref{tab3}.  These limits are more than
5 times smaller than the peak fluxes of the strongest single pulses from the
RRATs~\citep{mmclaugh2006, mmclaugh_iau}. For two RRATs, J0848$-$43
and J1839$-$01, the peak fluxes of the strongest pulses are 100~mJy at
1400~MHz, which means that they should be even larger at our frequency of
820~MHz, assuming typical pulsar spectral indices. If XDINSs
manifested the same single-pulse properties of the RRATs, they should
be easily detected in our observations at about the $60\sigma$ level.

\subsection{Test observations of known radio pulsars}

We successfully applied the same search pipeline to data taken
during the same observing session with the same observational
setup on the 246-ms
pulsar B0540+23, the relativistic binary pulsar B1534+12, and the
1.24-s pulsar J0628+09. They were detected with all three search
methods (see, for instance, Figs.~\ref{fig3} and
\ref{fig7}).  For the pulsar J0628+09 the first FFT run
did not detect the main pulsar harmonic using the default incoherent
foldings of 32 harmonics.  The subsequent runs with the summation of
64 and 128 harmonics found the pulsar at a frequency of 0.8055~Hz ($P
= 1241.477$~ms) with S/N $=17.2$ and $19.5$, respectively. The value
of 128 summed
harmonics corresponds roughly to the pulse width of about 10~ms
($N_\mathrm{harm} \sim P/w$). The S/N of the integrated profile 
with a bin width of 10~ms
is consistent with the spectral S/N and is about 20. This
corresponds to a mean flux density of about $40~\mu$Jy at
820~MHz. Note that this pulsar was
originally detected only through a single-pulse search  \citep{cordes_palfa2006}.
 For the relativistic binary
B1534+12 the FFT and FFA searches give S/Ns of about 11 and 8,
respectively. 
Including an acceleration search
increases the S/N to 30, corresponding to a mean flux
of about 0.35~mJy. 
This value is smaller than the expected mean flux at our
observing frequency of about 3.4~mJy \citep{wol1991,kramer1998}
and can be accounted for by scintillation.
The pulsar B0540+23 was found with S/N of 76 in the FFT search
with 32 harmonics folds and of about 240 in the FFA search.  The mean flux
of about 7~mJy that corresponds to S/N = 240 is about 3 times less
than the expected flux at this frequency. 
This can be due to interstellar scintillations with a timescale of about
several minutes (observing time was only 1 min). System overflow
could also have been an issue
because the pulsar is very strong, and the inherent 3-level
sampling of the SPIGOT is subject to saturation for bright sources.
The presence of
artifacts in the average profile further supports this suggestion.

\section{Discussion}\label{discussion}

The elusive radio emission of XDINSs, the sporadic 
radio activity
of the RRATs, and
the recently discovered transient radio emission from allegedly
`radio-quiet'' magnetars
 \citep{camilo2006,camilo2007} hint at close
relationships between these classes of neutron stars.
\citet{popov2006} have shown that the implied birthrate of RRATs is
more consistent with that of XDINSs than that of magnetars.  As shown
in the \ppdot diagram (Fig.~\ref{fig1}), RRATs and XDINSs also
have similar periods and period derivatives, implied ages and magnetic
fields.  
Moreover, X-ray observations of one RRAT~\citep{mmclaugh2007} reveal properties
similar to those of XDINSs (i.e.~a purely thermal spectrum 
with $kT\sim 140$~eV and a broad absorption feature at $\sim 1$~keV).
However, the RRATs spin-down properties are also consistent
with those of the normal pulsar population and transient magnetars in
quiescence. The nearby ($\simeq 300$~pc) radio pulsar
B0656+14, in addition to underlying weak emission of broad pulses,
manifests extremely bright spiky emission \citep{weltevrede2006}. 
Because of this,
it would only be detected as an RRAT were it
placed at a distance of a few kiloparsecs \citep{weltevrede2007}.

The distances to the XDINSs are also
believed to be much smaller than those to the RRATs ($\lesssim
1$~kpc). Thus, we should have had high sensitivity to RRAT-like radio
emission. Indeed, with the average $1\sigma$ sensitivity to single
radio pulses of 4~mJy for the XDINSs, the strongest pulses, with the same properties as those of
the RRATs \citep{mmclaugh_iau}, would be detected with S/N $> 60$. 
However, we have not detected such isolated radio pulses
or any periodic emission from the six XDINSs we have observed.
Our non-detection of such emission, however, does not necessarily mean that there is
no relationship between the two source classes.

The radio emission of
XDINSs may simply be more sporadic than that of the RRATs.
We estimated the upper limits
on the rates of possible pulses from the XDINSs as less than one event in the
full observing time for every source (see Table~\ref{tab3}, column~5). 
These limiting pulse rates range from 0.24 pulses per hour for RX J1308.6+2127 to
the maximum rate of 0.36 for RX~J2143.0+0654.
The smallest average pulse detection rate of the known RRATs
is 0.3 pulses per hour for J1911+00. 
This value corresponds exactly to the average upper limit on pulse detection rates
for all XDINSs. However, some RRATs, like J1839--01, show extreme 
variations in their pulse detection rates and have periods where
the detection rate is much less than 0.3 pulses per hour.

Another possibility is that XDINSs may be truly ``radio-weak'' or
``radio-quiet''. The derived upper limits for the radio luminosities of pulsed
emission at 1400~MHz $L_{1400}^\mathrm{p,max}$ are very small,
$(0.14-5)\times 10^{-3}$~mJy kpc$^2$
(Table~\ref{tab3}, column 3). No normal radio pulsars are known
to have such low values of radio luminosity at 1400~MHz, with only
six radio pulsars having values smaller than 0.1~mJy
kpc$^2$.\footnote{See ATNF pulsar catalog at
\url{http://www.atnf.csiro.au/research/pulsar/psrcat}.} However, this
could be due to our lack of knowledge about the spectral indices of
XDINSs. In our estimates we have used the mean value of spectral
indices for normal radio pulsars of $-1.8$. To avoid using such an
uncertain quantity we also compared radio luminosities of XDINSs with
that of normal radio pulsars at our frequency of 820~MHz. Upper limits of the radio
luminosities of XDINSs at 820~MHz are listed in column 4 of 
Table~\ref{tab3}. For comparison we chose 285 pulsars with
known spectral indices and radio luminosities at either 1400 or
400~MHz. The lowest derived value of radio luminosity at 820~MHz is
0.17~mJy kpc$^2$ for PSR J0030+0451 with only 8 pulsars having
luminosities below 1~mJy kpc$^2$. Thus, by any means the
pulsed radio luminosities of XDINSs are much weaker than those
of normal radio pulsars. 
Column 7 of Table~\ref{tab3} lists the upper limits
for single-pulse luminosity at 1400~MHz. They are all less than 10~mJy kpc$^2$,
but probably the true value is even less, $< 1$~mJy kpc$^2$, as for
RX J0720.4$-$3125 and RX J1856.5$-$3754 with measured parallax
distances.  The minimum single-pulse luminosities for RRATs are believed to be
about
100~mJy kpc$^2$ \citep{mmclaugh2006, mmclaugh_iau}.  
Therefore, the ``radio-weak'' or ``radio-quiet'' scenario is consistent
with our results.

Perhaps the most likely possibility, however, is that XDINSs simply
represent a sub-sample of long-period ordinary radio pulsars with
relatively high magnetic fields.  Being nearby objects, their radio
emission could escape detection in our deep observations due to
unfavorably aligned radio beams.  Indeed, assuming the radio pulsar
mean beaming fraction $f = 0.1$ \citep{tauris1998}, i.e.~the
probability that any pulsar's emission beam intersects our line of
sight, then the probability that all six XDINSs are beamed away from
the Earth is $(1 - f)^6 = 53\%$.  For long-period pulsars this
probability could be even as high as $83\%$, if instead of the mean
value of $0.1$ one uses the relation (15) from \citet{tauris1998}
between beaming fraction and the pulsar period, for the periods from
Table~\ref{tab2}. In other words, we would need a sample of about 40
sources to provide a $1\sigma$ probability of at least one of their
radio beams intersecting our line of sight.

Are there any long-period high-$B$ normal radio pulsars known to emit
in X-rays, and if so, are their properties consistent with those of
XDINSs? There are 34 rotation-powered radio pulsars in our Galaxy, listed in the ATNF pulsar catalog,  with
periods $> 1$~s and magnetic fields $B > 10^{13}$~G.
As discussed,
the X-ray properties of RRAT J1819$-$1458 are
remarkably similar to those of XDINSs \citep{mmclaugh2007}.
To our knowledge only four of the remaining 33 
--- B0154+61, J1718$-$3718, J1814$-$1744 and J1847$-$0130
~--- have been observed in X-rays. No X-ray emission was
detected from the $B = 2.1\times10^{13}$~G, $P$ = 2.3~s PSR~B0154+61
 in a 31-ks observation with XMM-{\em Newton} \citep{gklp04}.
Archival \emph{ROSAT} and \emph{ASCA} data 
(duration of 7.7 and $2\times 11.5$~ks, respectively) 
do not reveal X-ray emission from the field
of  the $B = 5.5\times10^{13}$~G, $P$ = 3.9~s  PSR~J1814$-$1744 \citep{pivovaroff2000}. A recent 6-ks
observation with \emph{Chandra} also resulted in a non-detection
(PI: Camilo, proposal number 0207010101). 
No X-ray emission was also found from the 6-ks
\emph{ASCA} archival data of the field containing J1847$-$0130, with
$B = 9.4\times10^{13}$ and $P$ = 6.7~s.
\citep{mmclaugh2003b}. However, for pulsar J1718$-$3718, with
$B = 7.4\times10^{13}$~G and $P$ = 3.3~s, very faint
X-ray emission was found by \citet{kaspi2005} in a much longer 55.7-ks
\emph{Chandra} observation. The thermal spectrum
 resembles that from
XDINSs, but a longer observation of J1718--3718 is needed to
draw any firm conclusions.

While B0154+61 has a DM-derived distance of only 1.6~kpc,
the non-detection of X-ray emission from the very high magnetic field pulsars J1847--0130 and J1814--1744 could
 be due to their large distances (7.7~kpc and
9.8~kpc, respectively, compared with 4.9~kpc for X-ray faint J1718$-$3718)
and the shorter integration times used. 
If deeper X-ray observations would allow to detect and recognize them as XDINSs, this 
will strongly support the hypothesis that XDINSs are simply nearby long-period
high-$B$ normal radio pulsars beamed away from us.
Rapidly evolving to
long periods, they still can be detected in X-rays if they are nearby
and cooling of the neutron star is not yet completed. Further
X-ray and optical observations with long integration times of
long-period high-$B$ radio pulsars are necessary to confirm their
relation with XDINSs.

This attractive picture, however, does not agree well with the results from
population syntheses. Under the standard assumptions about
neutron star cooling 
\citep[see][~and references therein]{page2006}
the birth rate of XDINSs is higher than that of
high-$B$ pulsars \citep{popov2006, vranesevic2004}.  The XDINSs can be
older, and so have a lower birth rate consistent with that of
long-period high-$B$ radio pulsars, only if their cooling does not
follow the standard model \citep[e.g.,][]{blaschke2004,page2004,yakovlev2004}.  This,
however, is very improbable because standard population synthesis
models reproduce the local population of neutron stars very well
\citep[e.g.,][]{popov2006b}. The assumption that the birth rate of
high-$B$ pulsars is underestimated, though possible, does not
correspond to the log-normal distribution of the magnetic fields of the pulsar
population \citep{fg2007}. A joint population synthesis of radio
pulsars and cooling isolated neutron stars 
with non-standard assumptions~---
such as, heating due to magnetic field decay and influence of magnetic fields on cooling \citep{pons2007, aguilera2008} 
--- could be very relevant.
It is certainly possible
that XDINSs represent a small group of sources with unique
properties and narrow radio beams which do not intersect our line of sight.

Searches at lower frequencies, where radio
emission beams are believed to be wider \citep{radha1969}, may be more sensitive to
radio emission from XDINSs. Indeed, 
\citet{malofeev2005, malofeev2007}
reported the detection of
radio emission from RX~J1308.6+2127 and RX~J2143.7+0654 at the low
frequency of 111~MHz. If the detection of Malofeev and co-authors is real,
our non-detection of radio emission from these two XDINSs at 820~MHz
implies that their spectral index $\alpha<-4$. 

In summary, we have presented the most sensitive limits so far on
periodic and transient radio emission from XDINSs. 
The lack of detection makes comparisons between
XDINSs and other radio populations of neutron stars difficult. It
is possible that the radio emission from XDINS is truly very weak,
but it is also
quite plausible that XDINs have normal radio luminosities, but are
simply beamed away from us. Observations
at lower frequencies are necessary to further constrain the
presence of any radio emission. 
Extensive high-energy studies (X-ray, optical) of known long-period high-$B$
normal radio pulsars and RRATs are required to support the hypothesis that they belong to the same group as XDINSs.
The detection of more XDINSs
and subsequent radio followup observations are also crucial in this regard.

\acknowledgements

Authors greatly thank the anonymous referee for her/his very valuable comments
that improved the paper considerably.
VIK also thanks Anna Bilous from the University of Virginia for useful discussions of
analytical approach of the FFT search.
VIK, MAM and DRL are supported by a Research Challenge Grant
from  WVEPSCoR. 
MB and AP have received support by the Ministero dell'Istruzione,
dell'Universit\`a e della Ricerca, under the national program PRIN-MIUR2005.
SBP thanks INTAS for financial support.
SZ thanks STFC (ex-PPARC) for support through an AF.
MB and RT are partially funded by INAF-ASI through grant AAE TH-58.
The Robert C. Byrd Green Bank Telescope (GBT) is
operated by the National Radio Astronomy Observatory which is a
facility of the U.S. National Science Foundation operated under 
cooperative agreement by Associated Universities, Inc.

\appendix
\section{Fast Folding Algorithm}\label{ffa}

The Fast Folding Algorithm (FFA) was originally developed by
\citet{staelin1969} to detect periodic signals in noisy data. While
the Fast Fourier Transform (FFT) accomplishes this task in the
frequency domain, the FFA works in the time domain.  The FFA 
is much faster than simply folding over a range of trial periods.
The computational time is reduced 
by avoiding the redundant summations
for many trial foldings for different periods. In this appendix we will
not repeat the mathematical details of the algorithm, but
will rather focus on specific aspects of applying this technique to
pulsar searches.  The necessary background on the FFA itself can be
found both in the original paper by \citet{staelin1969} as well as in
\citet{lovelace1969}, \citet{burns1969}, \citet{hankins_rickett1975}, and \citet{handbook}.

\subsection{Description}

We start with a time series of $N$ samples taken with sampling
interval $\Delta t$ which contains a periodic signal too weak to be
detected over the noise without folding. For pulsar searches, this
would be a time series dedispersed at some particular dispersion
measure, DM.  A single FFA transaction calculates the number of trial
folded profiles $M = N/P_0$ in the range from $P_0$ to $P_0 + 1$,
where $P_0$ is the trial base period in samples. The trial base period
in time units $P_0^\mathrm{t}$ is then $P_0^\mathrm{t} = P_0
\times\Delta t$, and the period of every folded profile can be
determined by
$$
P^\mathrm{t} = \left[P_0 + \frac{1}{M-1}i\right]\times \Delta t = P_0^\mathrm{t} + \frac{P_0^\mathrm{t}}{N-P_0}i~,
$$
where $0\leq i\leq (M-1)$\footnote{Note that the formula given in
\citet{staelin1969} is incorrect though it gives right values of
periods for the cases of $N=12$ and $P_0 = 3$.}.  The step
$\Delta P$ is determined by the integer number of 
periods in the time series and the sampling interval of the data, 
i.e.~$\Delta P = \Delta t / (M-1) = (P_0 \cdot \Delta t) / (N-P_0) =
P_0^\mathrm{t} / (N-P_0)$.  The only thing the FFA requires is for the
ratio $M$ to be a power of 2, or $\log_2 (N/P_0)$ to be an
integer. Another implementation of the original FFA by
\citet{lovelace1969} requires both $N$ and $P_0$ to be a power of 2
rather than just their ratio.  Without the FFA this procedure would
require $N^2/P_0$ summations, whereas with the FFA it is just $N\log_2
(N/P_0)$.

In practice, the usual period range to be searched can be as large as
several seconds, or in some specific cases only a few milliseconds (for
instance, when you know the period from observations at other
wavelengths, as in the case of the XDINSs).  The FFA search will
typically result in higher signal-to-noise ratio than FFT for pulsars with periods
$> 6$~s and pulse duty cycles up to 4\%, or even for pulsars with periods $\gtrsim 2$~s
and small duty cycles up to 0.8\% (see below).
In the frequency domain, these long periods correspond to
frequencies of 1~Hz and less, which are in the lower part of the
spectrum, presumably dominated by red noise, making the FFT less
efficient than the FFA.

\subsection{Implementation}

\begin{figure*}[tbph]
\includegraphics[angle=270,scale=0.65]{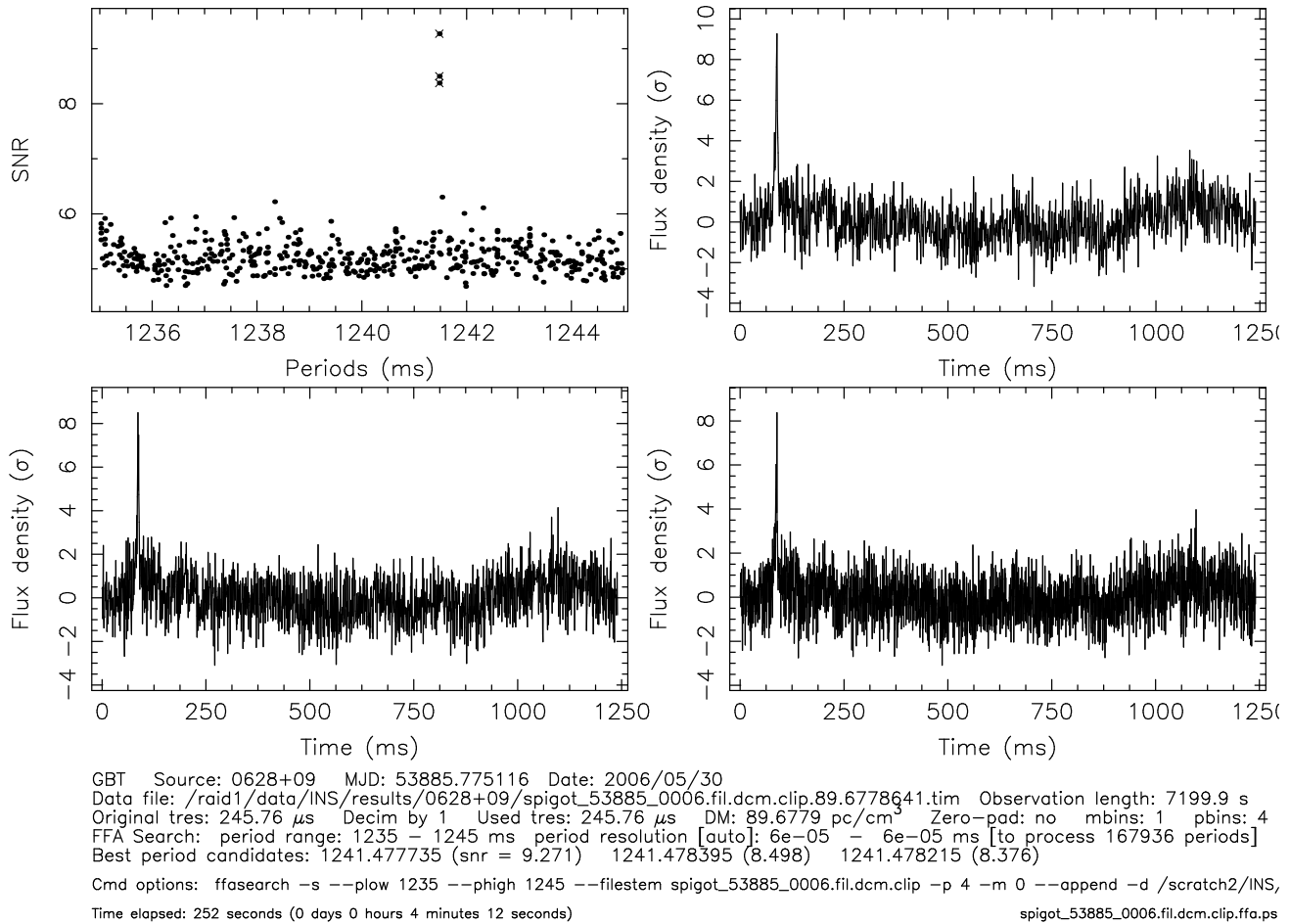}
\caption{FFA diagnostic plot for a 2-hr observation of the 1.24-s pulsar J0628+09.
The top left plot represents the periodogram, i.e. the S/N of the most
significant peak in the folded profile as a function of trial
period. The three crosses in the plot mark the S/Ns of the three best
candidates. The other three plots, top-right and from left to right on
the bottom, are the folded profiles for those three best
candidates. The profile of the best candidate (top-right) was folded
with a trial period very close to the true topocentric period of 
pulsar J0628+09 and has a S/N of about 9. 
The matched filtering technique was applied with the rebinning of folded profiles
by up to 4 samples ($983~\mu$s).
The legend at the bottom shows
the range of the searched periods and trial periods of the three best
candidates together with some other information.
}
\label{fig6}
\end{figure*}

The described FFA method was implemented as a C program
\emph{ffasearch}\footnote{The program can be downloaded from
{\url{http://astro.wvu.edu/projects/xdins}}.} to search for 
pulsars in noisy time series.
It executes successive single-algorithm
operations for different $P_0^\mathrm{t}$ from $P_\mathrm{low}$ to
$P_\mathrm{high}$, the values of which are specified in the command
line.  As the result, the sifted and full lists of candidates are
created together with a plot containing the periodogram and
plots of the folded profiles for the three best candidates.  An example
of this diagnostic plot is shown in Fig.~\ref{fig6} for the
1.24-s pulsar B0628+09.  Some details of the candidate selection
procedure, the candidate sifting, rebinning, and duty cycles  
issues are given briefly below.

\placefigure{fig6}

{\bf Candidate selection.} For every base trial period $P_0$ we have $M$
different folded profiles with periods spaced by $\Delta P$.  From
them we select only the three best candidates with the highest signal-to-noise
ratios (S/N) of the maximum peaks in their profiles, S/N = $(I_\mathrm{max}
- \langle I\rangle)/\sigma$, where the average intensity $\langle
I\rangle$ and rms $\sigma$ in the folded profile are calculated by
excluding a window centered on the peak $I_\mathrm{max}$. The size of
this window is 20\% of the size of the profile.
For every candidate, we save S/N, period, the phase of the maximum,
the number of other candidates with smaller
S/Ns with the same phases of the three best candidates, and duty cycle (see below). The full list of all detected candidates
is recorded into an ASCII file with other information such as DM,
sampling interval, etc.

{\bf Sifting.} Even with only three candidates for every trial base
period $P_0$, the final output list is very large, with many
candidates having similar periods.  We therefore sift the full list,
sorting by S/N and 
excluding candidates with periods within 3\% of the period of any of
already sifted candidates. In the sifted list for every candidate the
number of other candidates with close periods that were excluded is
noted.

{\bf Rebinning.} Often the sampling interval of the time series is
very small in comparison with the expected pulse width of a
pulsar. Therefore, rebinning will help to increase the S/N and speed
up the search process. There are three different types of rebinning
that were implemented: (a) the preliminary decimation of the time
series by several samples before starting the FFA; 
(b) rebinning of 
each of $M$ trial periods $P_0$ before starting the FFA search;
(c) extra rebinning of every trial folded profile after the FFA run.
Both (a) and (b)
decrease the time resolution of the original time series, increase
$\Delta P$ and decrease the execution time. 
However, the ``b''-rebinning  has the major advantage over the ``a''-decimation
that for every trial $P_0^\mathrm{t}$, 
all $M$ periods are summed in phase,
thus increasing the S/N of pulsar candidates (the ``a''-decimation and ``b''-rebinning
are the same only when $P_0$ is an integer multiple of the decimation factor, which is very unlikely).
The extra ``c''-rebinning represents the matched filtering technique described 
by \citet{cordes_mam2003} for every folded profile after FFA run and was implemented 
to be sensitive for pulsars with different pulse widths. 
This rebinning will not
decrease the period resolution $\Delta P$ and is a small increase
in the execution time in comparison with the time required to run
through all FFA operations.
For long time series, it is typically necessary to use the preliminary decimation to fit
the entire data segment into memory, and then use the ``b''-rebinning to ensure reasonable
execution times.

It is important to distinguish between rebinnings (b) and (c). The latter 
is applied to each of $M$ {\it trial folded profiles} for periods between 
$P_0^\mathrm{t}$ and $P_0^\mathrm{t} + \Delta t$ {\it after} the FFA. This procedure is the one
that was referred to by \citet{lovelace1969} as searching different pulse
widths ``by adding the sums corresponding to different phases''.
The ``b''-rebinning is applied not to trial folded profiles but to each of $M$ 
{\it periods} $P_0$ {\it before} the FFA run. This has the advantage of significantly decreasing
the number of FFA summations and the FFA execution time by reducing
the number of bins in the final folded profiles (if there is no need for smaller
time resolution).

{\bf Duty cycles.} The choice of rebinning determines the duty cycles
of the pulsars to which the FFA is most sensitive. 
Owing to the definition of S/N based on the single maximum bin in the
folded profile, optimal detection will occur when the pulse width 
is similar to the final sampling interval.
Therefore, one should consider the overall
preliminary decimation and rebinning within the FFA as the one that
determines the minimal pulse width (or duty cycle) optimized in this particular
search. Thus, if one searches the data from $P_\mathrm{low}$ to
$P_\mathrm{high}$ with an interval between adjacent bins of $\Delta
t_\mathrm{ffa}$ then duty cycles ranging from $\Delta
t_\mathrm{ffa}/P_\mathrm{high}$ to $\Delta
t_\mathrm{ffa}/P_\mathrm{low}$ are covered optimally.
 However, to increase the sensitivity of the search to
pulsars with larger duty cycles, it is worthwhile to rebin all the
folded profiles by different factors covering a larger range of duty
cycles \citep[i.e. the matched filtering technique, see][]{cordes_mam2003}.
The \emph{ffasearch} program is now implemented in a way to
perform 
this matched filtering (or ``c''-rebinning)
of folded profiles after the FFA run for a
selected range of pulse widths with an increment of 1 bin.

\begin{figure}[tbp]
\includegraphics[angle=270,scale=0.65]{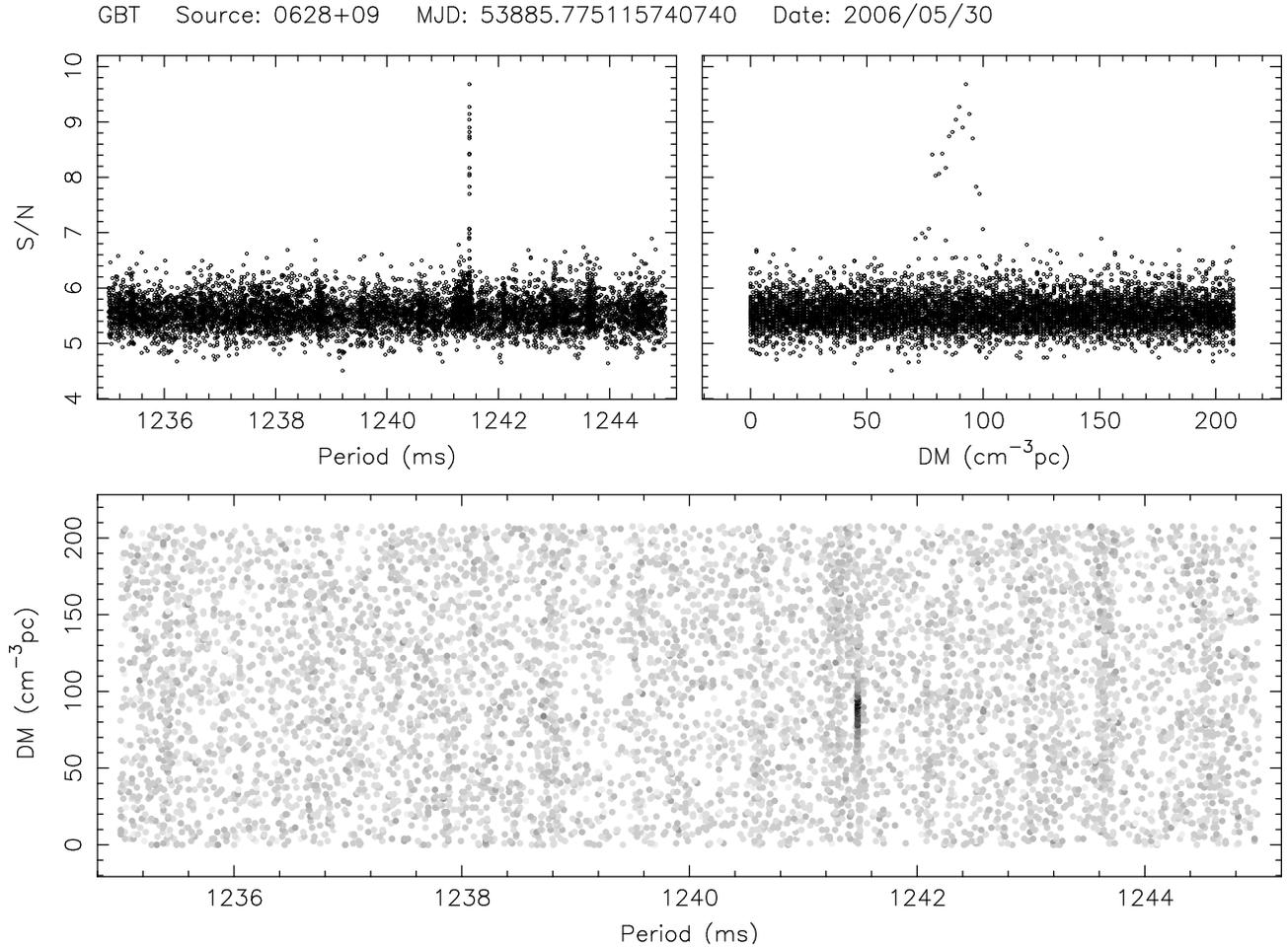}
\caption{
An FFA diagnostic plot combining the results for a range of trial DM
values for the 1.24-s pulsar J0628+09.  The bottom plot represents
DM-$P$, with darker points corresponding to candidates with higher
S/N. The two top plots are S/N of the candidates as a function of a
trial period $P$ (top-left) and trial DM (top-right). The actual
pulsar values of period and dispersion measure are clearly seen in all
three plots.
}
\label{fig7}
\end{figure}

In addition to implementation of the FFA itself,
a \emph{Perl}-script, \emph{ffadmplot}, was
written that combines the FFA results for all trial values of DM in a
number of diagnostic plots. An example of this plot for pulsar
J0628+09 is shown in Fig.~\ref{fig7}.
At the bottom is  a DM-$P$ gray-scale plot with darker points corresponding
to candidates with higher S/Ns.
The other two plots on the top show S/N as a function
of trial period and DM. For the bottom plot, one can use either grayscale
or circles such as used for the single-pulse diagnostic 
plots\footnote{See \url{http://astro.wvu.edu/projects/xdins} or 
examples of single-pulse diagnostic plots in \citet{cordes_mam2003}.}. 
This
script has interactive plotting, allowing the user to examine every point
in all of three plots, select the candidates, and make a plot with
folded profile for every selected profile. As shown in
Fig.~\ref{fig7} the true pulsar period is clearly seen
in the bottom plot.

\placefigure{fig7}

\subsection{FFA vs. FFT}\label{simul_section}

\begin{figure}[tbp]
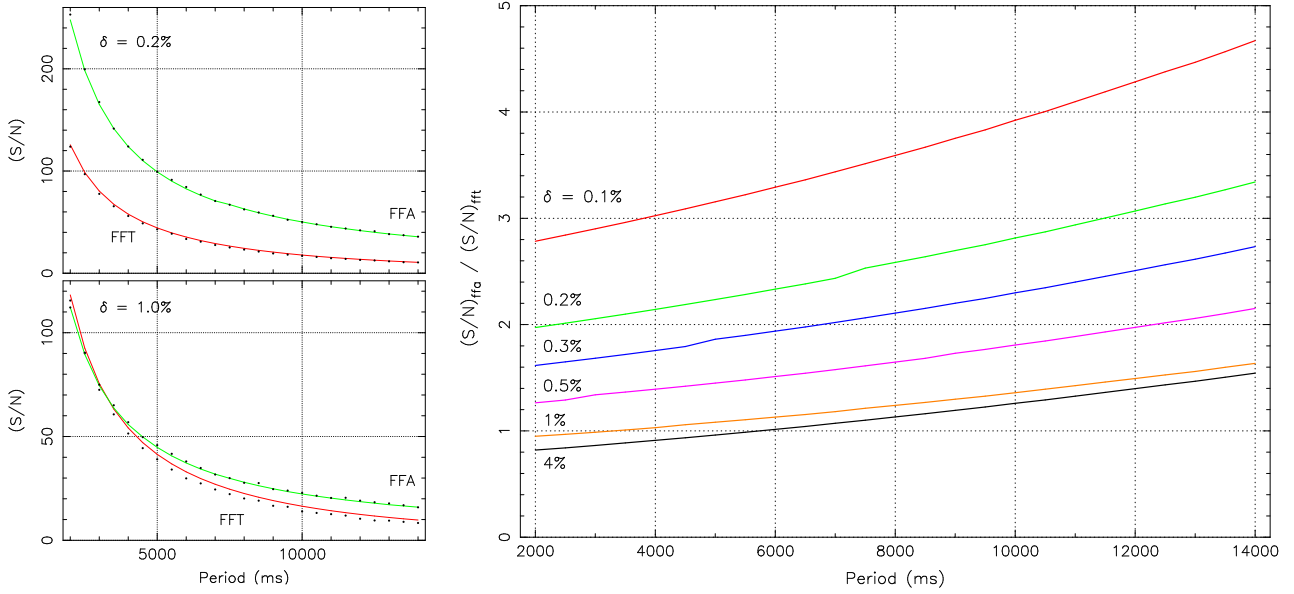

\begin{flushleft}
%
%
%
\hskip 5mm\includegraphics[angle=270,scale=0.21]{f8a.eps} \\
\hskip 5mm\includegraphics[angle=270,scale=0.21]{f8b.eps}\vskip -7.8cm\hskip 6.5cm\includegraphics[angle=270,scale=0.42]{f8c.eps}
%
%
%
\caption{
{\bf Left.} The signal-to-noise ratio (S/N) versus the pulsar period for two different duty cycles of $\delta=0.2\%$
(top) and $\delta=1\%$ (bottom) for both FFT and FFA searches. Results from our simulation are plotted by dots, and
the dependences of S/Ns from the analytical consideration are shown by lines. {\bf Right.} The ``efficiency'' 
of the FFA search versus the pulsar period and as a function of duty cycle. The ``efficiency'' is defined as the
ratio of the S/N of the fake pulsar detected with the FFA search to the S/N of the fake pulsar detected with 
the FFT search. Six different curves represent the analytical dependence for different pulse duty cycles.
\label{fig8}
}
\end{flushleft}
\end{figure}

{\bf Simulation.} To compare the efficiency of both the FFA and the FFT, we 
performed FFA and FFT searches on artificial pulsar signals with different periods 
and pulse widths. For this purpose we used the \emph{fake} program from the \emph{SIGPROC}
package. This program creates time series of Gaussian noise with injected top-hat pulses with
predetermined period, width and peak signal-to-noise ratio.
Then we ran both FFT and FFA searches on these
time series with the fake pulsar, recording the S/N with which the pulsar was detected in both searches.
The sampling interval was chosen to be $491.52~\mu$s, the same as in the real XDINS data. The number
of generated samples corresponded to an observing time of 1 hour. The simulation was done
for pulsar periods, $P$, ranging from 2 to 14~s with a step of 0.5~s and for 14 different values of pulse duty
cycle, $\delta$, covering $0.1-1$\% (with stepsize 0.1\%) and $2-5$\% (with stepsize 1\%).
 Therefore, in total we have 350
fake pulsars with different $P$ and $\delta$. The peak fluxes of injected top-hat pulses were different
for different $P$ and $\delta$ to keep the total pulse energy the same among all trial fake pulsars. Thus, the
peak flux density was less for broader pulses. 
The number of harmonics summed in the FFT search was 32.
For every combination of $P$ and $\delta$ we performed 30 different trials recording the S/Ns from both
searches. The results of our simulations are shown on Fig.~\ref{fig8} (left, dots) for two cases 
of $\delta = 0.2$ and $1\%$. For other duty cycles the obtained dependencies are similar.

The plots show the dependence of the S/N for both FFT and FFA searches as a function of period. In addition
to the results from the simulation, the lines represent the derived analytical dependence (see below). 
The apparent decrease of the S/Ns with increasing period and duty cycles is due to 
our requirement of constant energy for all individual pulses, and because of fewer 
numbers of larger periods in our fixed observing time.
It is clearly seen that for smaller duty cycles (0.2\%, narrower profiles) the FFA search performs better 
than the FFT even for small periods. However, this is not that relevant as most pulsars with duty 
cycles $< 1$\% have long periods. For instance, of 25 radio pulsars in the ATNF catalog 
with duty cycles $<0.8$\%, 19 have period $>1$~s. The broader the pulse (the larger $\delta$), 
the longer the pulsar period should be in order for the FFA search to be more efficient than the FFT search,
particularly for the $\delta = 1\%$ the FFA search is only more  efficient than FFT search for
periods more than 4~s.
It is worth mentioning that fake pulsars with periods $>10$~s and large duty cycles of 2--5\% 
were not detected in the FFT search in \emph{every} trial. 
This demonstrates the importance of using an FFA search for very long period pulsars.


{\bf Analytic approach.} In addition to the simulation, we can also make a comparison between the
FFA and FFT effectiveness from analytical considerations. Our simulation showed that the
FFA is advantageous for pulsars with short duty cycles. This is because short duty cycle
pulsars distribute their power over many harmonics, and only a limited number
of harmonics are summed together in the FFT search. This number depends on the particular FFT
search procedure used, but a typical maximum number of summed harmonics is 32 (as in the 
current search and simulation).
Moreover, these harmonics are summed {\it incoherently}, further  reducing the S/N
that can be achieved with the FFT. Contrary to the FFT search, the FFA search folds all the data in phase,
or {\it coherently}.
Larger S/Ns can also be reached for pulsars with larger duty cycles (broader pulses)  by rebinning 
the profiles with an optimal window equal to pulse width.
Thus, assuming that no power is lost in the FFA search, and estimating the power 
that can be recovered 
during an
incoherent sum of a finite number of harmonics for the FFT search, we can derive an analytical
expression for the ratio of detected S/Ns for the FFA and FFT search for a particular pulsar.

Consider for simplicity the periodic train $\Pi(t)$ of top-hat pulses with amplitude $S$, period $P$,
and pulse width $w$ in the presence of noise $n(t)$ with zero mean and standard deviation
of unity. 
We can write the pulse amplitude $S$ as $E_p/\delta P$, where $\delta = w/P$ is the pulse
duty cycle, and $E_p = wS$ is the pulse energy, chosen to be constant for all artificial
pulsars in our simulations described above. The signal-to-noise ratio, 
$(\mathrm{S/N})_\mathrm{ffa}$, of the single pulse
in a non-decimated time series $\Pi(t) + n(t)$ is then equal to its amplitude $S$. By folding all periods
together we increase the $(\mathrm{S/N})_\mathrm{ffa}$ by $\sqrt{N_p}$, 
where $N_p = N/P = T/P^t$ is the total number of periods in the time series, $N$ the total 
number of samples, $T=N\Delta t$ the observing time (1 hour in our simulation), $\Delta t$ the
sampling interval ($491.52~\mu$s in the simulation) and $P$ or $P^t$ the pulsar period either in
samples or time units. One can also increase the $(\mathrm{S/N})_\mathrm{ffa}$ by using the matched
filtering technique, i.e. by rebinning the folded profile with the optimal window of pulse
width $w$. This will increase $(\mathrm{S/N})_\mathrm{ffa}$ by $\sqrt{w}$.
Thus, finally, the signal-to-noise ratio of our pulse train $\Pi(t)$ from the FFA search becomes
$$
(\mathrm{S/N})_\mathrm{ffa} = \frac{E_p\sqrt{N}}{P^t\sqrt{\delta}}~.
$$

To obtain the corresponding signal-to-noise ratio of the FFT-search $(\mathrm{S/N})_\mathrm{fft}$,
we derive first the Fourier response of our pulse train 
$\Pi(t) = \sum_{l=0}^{\infty} S\Pi\left[(t - lP + P/2)/w\right]$, where $\Pi(t)$ represents
the rectangular top-hat function of width and height equal to unity. The term $P/2$ reflects
the shift by half of the period from the origin (as in our simulated pulse trains).
It is easy to show that the complex $n$-th Fourier harmonic $F_n$ of this pulse train $\Pi(t)$ 
is given then by $F_n = -2 S (1-\delta) \mathrm{sinc}[\pi n(1-\delta)]$ 
\citep[see, e.g.,][]{bracewell2000}. The response of the discrete Fourier transform (DFT) $A_k$ in the 
FFT search for $n$-th harmonic is $NF_n/2$ for the ideal case when the frequency of the
harmonic coincides exactly with the frequency of the corresponding Fourier bin $k$, i.e. 
$nN/P = k$, $k = 0, 1, \ldots, N-1$. Usually this is not
the case and the harmonic amplitudes will be reduced by a factor of 
$\mathrm{sinc}[\pi(k-nr)]$, where $r$ is the real wavenumber 
defined as $r = N/P$ \citep[so-called ``scalloping'' effect; see, e.g.,][]{ransom2002}. 
However, in our FFT search the interbinning technique
was used that improves the DFT response with a loss of sensitivity of not more
than $\sim 7.4\%$. Therefore, in our further consideration we will only consider the case without
scalloping.

To estimate the average response of our harmonics in presence of noise and convert it 
into the S/N units we need to know
what are the values of the mean and root-mean-square deviation in our amplitide spectrum
$A_k = [{(c_k + a'_k)^2 + (d_k + b'_k)^2}]^{1/2}$, where $c_k$ and $d_k$ are real 
and imaginary parts of the spectrum related
to the signal, and $a'_k$ and
$b'_k$ are real and imaginary parts of the spectrum related to noise. The
$a'_k$ and $b'_k$ are independent and normally distributed with the mean zero and variance of $N/2$.
The normalized amplitude spectrum of such a kind obeys the non-central $\chi$-distribution
with two degrees of freedom \citep{evans2000, johnson1994}, and characteristic parameter 
${\lambda}_k = [(2/N)\cdot(c_k^2 + d_k^2)]^{1/2}$, where $c_k^2 + d_k^2 = N^2F_n^2/4$ for all
$k = N/P, 2N/P, \ldots$ and zero otherwise.
The mean, $\mu_k$ and rms, $\sigma_k$ of this
distribution are defined as $\mu_k = (\pi/2)^{1/2}\cdot L^0_{1/2}(-\lambda_k^2/2)$ and 
$\sigma_k = (2 + \lambda_k^2 - \mu_k^2)^{1/2}$, where $L^0_{1/2}(x)$ is the generalized
Laguerre polynomial $L^{(\alpha)}_n(x)$ of degree $n = 1/2$ and $\alpha = 0$. Hence, every
Fourier bin of our spectrum will have its own mean and rms values. 
For the harmonics related to our periodic signal $\Pi(t)$ in presence of noise 
the mean amplitude 
$\langle A_n\rangle$
will be equal to $(\pi/2)^{1/2}\cdot L^0_{1/2}(-N[S(1-\delta)\mathrm{sinc}[\pi n(1-\delta)]]^2)$.
To convert $\langle A_n\rangle$ into S/N units one needs to calculate the average $\langle\mu\rangle$ 
of $\mu_k\!$'s and average $\langle\sigma\rangle$ of $\sigma_k\!$'s
over the whole spectrum excluding Fourier bins related to signal. This corresponds to the case
with values of $c_k$ and $d_k$ equals to zero, and Laguerre polynomial $L^0_{1/2}(0) = 1$.
Thus, we have $\langle\mu\rangle = (\pi/2)^{1/2}$ and $\langle\sigma\rangle = (2 - \pi/2)^{1/2}$.
Finally, summing up the 32 harmonics in our FFT search, the FFT-search S/N of our 
artificial pulse train is
$$
(\mathrm{S/N})_\mathrm{fft} = \frac{1}{\langle\sigma\rangle\sqrt{32}} \sum\limits_{n=1}^{32}[\langle A_n\rangle - \langle\mu\rangle] = \frac{1}{4}\sqrt{\frac{\pi/2}{4-\pi}} \sum\limits_{n=1}^{32} [L^0_{1/2}(-N[S(1-\delta)\mathrm{sinc}[\pi n(1-\delta)]]^2) - 1]~.
$$

These analytical dependences of the $(\mathrm{S/N})_\mathrm{ffa}$ 
and $(\mathrm{S/N})_\mathrm{fft}$ are shown on the Fig.~\ref{fig8}, left, for two different duty cycles
of $\delta = 0.2$ and $1\%$. They demonstrate quite a good agreement with our simulation results,
especially for the FFA-search. The FFT analytical curve for the duty cycle of $1\%$ goes slightly
above the simulation data points, reflecting that we considered the ideal case without scalloping.
Also we used the theoretical values of the mean $\langle\mu\rangle$ and $\langle\sigma\rangle$
in the spectrum without the presence of signal. In real simulations these values are calculated
in windows excluding only strong outbursts of $> 3\langle\sigma\rangle$. However, the small deviation of
the argument in Laguerre polynomial can change the result noticeably.

In Fig.~\ref{fig8}, right we present the ``efficiency'' of the FFA search from the analytical derivation, 
or the ratio of S/N of the fake pulsar detected with the FFA search to the S/N of the fake pulsar detected
with the FFT search for a number of duty cycles. As was expected, for the longest periods, above 6~s, 
the FFA search gives larger S/Ns than the FFT for pulse duty cycles as large as 4\%. 
Due to slightly over-estimated S/Ns for the FFT search of our analytical expression, one should consider
the curves as a lower limit only. The divergency between the analytical value and that from the simulation
is not more than 5\% for duty cycle $\delta = 0.1\%$ and increasing for larger duty cycles (of about 20\%
for $\delta = 4\%$).


In practice, the efficiency
of the FFA should be even better if low-frequency noise is present.
We have checked our simulations using relatively RFI-free segments of
our XDINSs data
with injected fake pulses. Surprisingly, we did not find any significant difference between
results between the modeled Gaussian noise and real noise. This, however, can be strongly dependent on the
observational equipment.

In a blind search for pulsars with periods less than about 6~s the FFT search is preferable, assuming that the number
of possible pulsars with small duty cycles is low, and because the execution time is typically 
much smaller than in case of the FFA. 
However, for periods $>6$~s the FFA search is preferable to the FFT for a large range of pulse duty cycles.
Thus, the FFA search should be included in all searches for long-period pulsars.

\placefigure{fig8}

\subsection{Testing}

We have tested our program on a number of normal pulsars and on the two 
RRATs (J0848--43 and J1754--30) which have recently been shown to be detectable through
their time-averaged emission
in observations with the GBT at 350~MHz \citep{mmclaugh_iau}. 
These RRATs provide excellent
tests of the FFA because they generally have rather long periods and small duty cycles
that are similar to those of the XDINS.
 Both J0848--43 and J1754--30 were detected using both the FFA and FFT
searches with S/Ns that agree well with the FFA efficiency plot shown in Fig.~\ref{fig8}, right.  
Pulsar J0848$-$43
has a period of about 5.9~s and a pulse duty cycle of about 2\%, which should
result in roughly similar S/N detections in the FFA and FFT
based on Fig.~\ref{fig8}, right. Indeed, it was detected with S/N $\approx 16$ 
in the FFA search and with S/N $\approx 19$ in the FFT search.
Pulsar J1754$-$30
has a period of 1.32~s and duty cycle of about 8\%. As expected, it was
detected with larger S/N ($\approx 44$) in the FFT search than in the FFA search ($\approx 25$).
This agrees very well with our simulations that shows that the FFT search performs better for shorter
periods and broader pulses.

Tests on a few other pulsars confirmed this result. As an example, 
PSR J0628+09, with period of 1.24~s and small duty cycle of only 0.8\%,  was not detected in the FFT search at all using the default incoherent sum of 32
harmonics. However, it was detected with 64 and 128 harmonics with 
S/N of 17.2 and 19.5, respectively, owing to increased sensitivity to narrower pulses.
With the FFA it was even detected
without the matched filtering technique,
i.e. without optimal rebinning with smoothing window equal to the width of the pulse.
Both in the FFT and FFA, the S/Ns are about the same for
similar equivalent pulse widths, agreeing well with our simulation and analytical expression (see Fig.~\ref{fig8}).
The non-detection
of the pulsar with the FFT using the default incoherent sum of 32 harmonics is likely 
due to its weakness because of the sporadic nature.
This pulsar was discovered originally only in the single-pulse
search \citep{cordes_palfa2006}. Thus, in case of weak pulsars or bright but intermittent pulsars
the FFA has a significant advantage against the FFT.

\bibliographystyle{apj}
\bibliography{xdins}

\end{document}